# Long range pure magnon spin diffusion observed in a non-local spin-Seebeck geometry

Brandon L. Giles[1], Zihao Yang[2], John S. Jamison[1], and Roberto C. Myers[1-3]

[1]*Department of Materials Science and Engineering, The Ohio State University, Columbus, OH, 43210, USA*

Email: myers.1079@osu.edu , Web site: http://myersgroup.engineering.osu.edu

[2]*Department of Electrical and Computer Engineering, The Ohio State University, Columbus, OH, 43210, USA*

[3] *Department of Physics, The Ohio State University, Columbus, OH, USA*

The spin diffusion length for thermally excited magnon spins is measured by utilizing a non-local spin-Seebeck effect measurement. In a bulk single crystal of yttrium iron garnet (YIG) a focused laser thermally excites magnon spins. The spins diffuse laterally and are sampled using a Pt inverse spin Hall effect detector. Thermal transport modeling and temperature dependent measurements demonstrate the absence of spurious temperature gradients beneath the Pt detector and confirm the non-local nature of the experimental geometry. Remarkably, we find that thermally excited magnon spins in YIG travel over 120 µm at 23 K, indicating that they are robust against inelastic scattering. The spin diffusion length is found to be at least 47 µm and as high as 73 µm at 23 K in YIG, while at room temperature it drops to less than 10 µm. Based on this long spin diffusion length, we envision the development of thermally powered spintronic devices based on electrically insulating, but spin conducting materials.

# I. INTRODUCTION

Spin currents consist of a flux of angular momentum [1]. In magnetically ordered insulators, spin currents are carried by magnons. These spin currents are typically generated via the spin Hall effect [2] or through spin pumping, in which a microwave magnetic field excites spin-waves of approximately 10 GHz [3] . Such microwave frequency spin-waves have been proposed for use in logic devices with low power consumption [4], and recently an all-magnon based transistor was demonstrated [5]. The de-phasing length of such coherent spin-waves has been measured to be 31 μm at room temperature in YIG [6].

Recently, the spin Seebeck effect (SSE) [7–10] has been shown to provide an alternative method for spin excitation in magnetic insulators. An induced temperature gradient produces thermally-excited incoherent magnons that have much higher energies (500-6000 GHz) than the coherent microwave frequency magnons previously discussed (1-10 GHz) [11]. It is expected that thermally-generated magnons will couple more strongly to the lattice, and have shorter lifetimes and diffusion lengths than the microwave generated spin-waves. Nevertheless, these spin currents have stimulated a great interest within the field of spin caloritronics [12]. Additionally, they may result in the discovery of fundamentally new physics, such as the Bose-Einstein condensation (BEC) of superfluid magnons at the Pt/YIG interface [13]. However, the realization of such advances is hampered by the lack of quantitative measurements of the length scale over which these thermally-excited magnons exist.

In a recently reported thickness dependent measurement of SSE, Kehlberger *et al*. suggest a thermal magnon spin diffusion length of approximately 100 nm at room temperature [14]. The observed saturation of the SSE signal for films thicker than 100 nm was interpreted to indicate the length scale over which thermal magnons could contribute to the signal. Theoretically, this length scale was predicted to be 70 nm at 300 K, based on modeling of the SSE in YIG/Pt [15], which is consistent with this observation. Kehlberger *et al*. also find that the thermally excited magnon propagation length reaches ~ 7 µm at 50 K. However, the strong interface sensitivity of the SSE leads to measurement uncertainty due to sample variation i.e each Pt/YIG interface is different between different samples [16]. Additionally, in the longitudinal SSE geometry, the applied heat flows directly into the spin detector, allowing for the possibility of contamination of the measured inverse spin hall signal from other charge-based thermo-electric effects [17] such as the anomalous Nernst effect (ANE). This motivates us to design a new experiment capable of probing the diffusion length of thermally-excited magnon spins outside of the vicinity of any thermal gradients while simultaneously controlling for interface sensitivity, hence making a measurement that is immune to either parasitic thermoelectric effects or interface sensitivity.

Here we present systematic measurements of thermally-excited magnon spin diffusion in YIG, by utilizing a non-local detection geometry to sample spin currents induced by SSE. In metals and semiconductors, a similar non-local geometry was successful in measuring pure electron spin currents diffusing in the absence of parasitic effects caused by an applied electric field [18]. Because spin diffusion in the non-local measurements is caused solely by the gradient in chemical potential and not by driving forces, pure spin currents are directly isolated in such measurements, thereby providing a clean measurement of the diffusional spin flow in solids. By utilizing this non-local spin injection/detection geometry, pure magnonic spin currents are similarly isolated from the parasitic effects associated with an applied thermal gradient. This allows for an experimental determination of the magnon spin diffusion length in YIG. A similar non-local magnon spin measurement in YIG was very recently reported by Cornelissen *et al*. [19], where they utilized the spin Hall effect to inject a spin current into YIG and measured a spin diffusion length of 9 µm at room temperature. In the current experiment, we extend those measurements to 23 K and find a remarkable increase in the spin diffusion length up to 47 μm upon cooling the sample. The pure magnon spin current signal is observed even beyond 120 μm. Our measurements clearly show that, just as in the case of electron spin diffusion, magnon spins are preserved over many inelastic scattering events.

## II. NON-LOCAL SPIN-SEEBECK

### A. Experimental Setup

Thermal spin injection and detection is performed utilizing the opto-thermal method [20], in which a focused laser is absorbed by a Pt film that has been e-beam evaporated onto a bulk single crystal of YIG. A laser is modulated with an optical chopper and the resulting transverse voltage is sampled using a lock-in amplifier. Instead of using a continuous Pt film to detect the spin current, we pattern the film into a central Pt spin detector and an array of electrically and spatially isolated Pt absorption pads, as shown schematically in Fig. 1(a). When the laser is focused on an absorption pad, a local hot spot is produced, generating a thermal gradient ($\nabla T$) whose magnitude is varied by adjusting the average power of the beam. $\nabla T$ excites magnons in the YIG beneath the absorption pad, producing a spin current. Magnon spins that have laterally diffused underneath the spin detector can then transfer their spin via spin pumping into the spin detector, where they are converted into an electron spin current ($J^s{}_z$), which is then converted into a transverse voltage along the y-direction ($V_y$) by the inverse spin Hall effect (ISHE) [7] . The magnitude of $V_y$ is proportional to the magnitude of the spin current and its sign tracks the sign of the spin polarization injected into the detector, thereby following the sign of the in-plane (xy-plane) magnetization (***M***) of YIG.

### B. Thermal Transport in Non-Local Configuration

To ensure that the detected spins were generated non-locally, no significant heat flow can occur across the interface between the Pt spin detector and the YIG layer beneath. This is confirmed by using a finite-element method (FEM) to simulate heat flow and steady-state temperature gradient profiles caused by the heating laser. When the laser is focused on the Pt absorption pad, there are thermal gradients in both the vertical ($\nabla T_z$) and lateral ($\nabla T_x$) directions. FEM simulations of $\nabla T_z$ and $\nabla T_x$ directly below the heating laser are plotted in Fig. 1(b) and Fig. 1(c). The average laser power is 29.9 mW and the spot diameter is 7.5 μm. The simulations are for 23 K. Clearly, in the non-local configuration, heat flows from the absorption pad into the YIG. If the absorption pad is close enough to the detection pad, lateral heat flow through the YIG may lead to $\nabla T_x$ beneath the detector. However, because the detector is a thermal open circuit, the laterally diffusing heat cannot then diffuse vertically into the detector (see Appendix A). FEM modeling verifies this fundamental point, showing that $\nabla T_z$ is not present at the spin-detector. This verifies the non-local detection principle that the sampled spins are not generated at the detector/YIG interface as in longitudinal SSE [9]. FEM modeling further confirms that lateral heat flow, as indicated by $\nabla T_x$, decays to 1% of its peak value within 22.9 ± 1.08 μm. Therefore, when the laser is positioned at a distance greater than 23 μm from the spin detector neither substantial vertical, nor lateral, heat flow exists beneath the Pt detector. Any $V_y$ is attributable only to the diffusion of a pure magnonic spin current thermally produced at the remote Pt absorber. As will be shown in the following section and in Appendix B, even when the laser is closer than 23 μm the effects of $\nabla T_x$ are not present.

### III. RESULTS AND DISCUSSION

In Fig. 2, the raw transverse voltage across the detection pad, $V_y$, is shown as a function of an in-plane magnetic field ($B_x$) for measurements carried out in both the local and non-local configurations. In the local configuration, the laser is positioned directly on the Pt detection strip. As these spins are both generated and detected in the presence of $\nabla T_z$, $V_y$ may contain contributions from both the longitudinal SSE [9] and the anomalous Nernst effect (ANE) [21,22] . A schematic of the local configuration is shown in Fig. 2(a), and the corresponding $V_y$, measured across the Pt detection strip as a function of $B_x$, is presented in Fig. 2(b). $V_y$ saturates to 746 ± 3 nV at >|50| mT, which matches the saturation field of YIG, and the sharp $V_y$ hysteresis is consistent with the small coercive field of YIG. The M-H curve for this sample is included in Fig. 2(b) for comparison. We define the magnitude of the local voltage $V_L$ as the magnitude of the hysteresis loop. The inset plots $V_L$ over a range of laser powers, revealing a linear relationship consistent with both the SSE and ANE dependence on $\nabla T_z$. Note that in the absence of additional measurements it is not possible in the local configuration to determine the relative contributions of the SSE and the ANE to $V_L$ [23].

In the non-local configuration, the laser is positioned on one of the electrically isolated Pt absorption pads, as depicted in Fig. 2(c). Since $\nabla T_z = 0$, ANE cannot contribute to $V_{NL}$. The non-local configuration detects only the thermally-induced magnon spins that have diffused beneath the detector. Fig. 2(d) shows the corresponding $V_y$ in the Pt detection strip as a function of $B_x$, if the laser is on an absorption pad that is 54 μm from the detection pad. Again, $V_y$ tracks the magnetic hysteresis of YIG. $V_{NL}$ is defined similarly to $V_L$, as the magnitude of the measured hysteresis loop in the non-local configuration, but no longer contains any possible ANE contribution. It scales linearly with laser power, however its overall intensity is an order of magnitude smaller than $V_L$. Importantly, $V_{NL}$ also exhibits the same sign as $V_L$. This experimentally confirms that there is no accidental $\nabla T_z$ across the detector. If $V_{NL}$ were due to accidental heat flow from the YIG back into the detector, then $\nabla T_z$, and therefore $V_{NL}$ would change sign when the laser was moved from the detector to the absorber pad. This possibility is ruled out on both fundamental grounds, as described in Section II, and on experimental grounds (lack of observed sign change).

We also consider the possibility that parasitic heat flow could occur through the spin detector by thermal conduction through the detector wires leading to local magnon generation at the detector/YIG interface and contamination of $V_{NL}$. To generate the observed $V_{NL}$ would require use of unphysical thermal conditions (see Appendix A) and 0.4-mm thick wires. Since our measurement wires are 25 μm in diameter and the sample is held in high vacuum ($1\times10^{-7}$ torr), such parasitic thermal leakage can be ruled out.

To spatially map the thermally-generated magnon spin current, $V_{NL}$ is extracted from raw hysteresis traces (as in Fig. 2) with the laser positioned above absorption pads at increasing distances from the spin detector (Δx). Fig. 3(a) plots $V_{NL}$ extracted from such measurements as a function of Δx at 23 K. The same measurement is carried out at 280 K and plotted in Fig. 3(b). Additionally, the modeled $\nabla T_x$ at both temperatures are plotted in Fig. 3 to compare the spin transport versus the thermal transport. Since non-negligible $\nabla T_x$ is predicted near the edge of the spin detector, then in this region magnon spin transport may be driven by $\nabla T_x$ as in transverse SSE [7,8]. However, this is ruled out by comparing the measurements at 23 K to those at 280 K. At low temperature, $\nabla T_x$ decays much more rapidly than $V_{NL}$, whereas at 280 K, $V_{NL}$ is negligible at all values of Δx. Conversely, $\nabla T_x$ at high temperature exhibits a much larger magnitude than at low temperature due to the decrease in thermal conductivity of Pt and YIG at high temperature, however no $V_{NL}$ is observed at these conditions. The lack of $V_{NL}$ at high temperature is attributable to the reduction of the magnon spin diffusion length, which agrees with a recent report of room temperature measurement of magnon spin diffusion in YIG of 9 μm [19], which is smaller than the spatial resolution of our current experiment.

The $V_{NL}$ data are fit to a single decaying exponential over various ranges. The spin current decays exponentially with distance, consistent with the solution to the spin-diffusion equation at steady-state. From the exponential fits to $V_{NL}(\Delta x) = V_{NL}^s e^{-\Delta x/\lambda_S^*}$ we obtain the effective magnon spin diffusion length $\lambda_S^*$. Fig. 4 plots the results of non-local measurements carried out at three different laser powers, together with the exponential fits. $\lambda_S^*$ exhibits an average value of 47 μm independent of laser power. The lack of a change in $\lambda_S^*$ with laser power implies that the magnon spin scattering and lifetime are insensitive to the applied temperature gradient. There is no significant variation for fits to data sets that include data points gathered with the laser focused on differing sets of absorption pads (see Appendix B). Additionally, $\nabla T_x$ decays much more rapidly than $V_{NL}$ (see Appendix A). This analysis indicates that lateral heat flow has no detectable impact on the $\lambda_S^*$ measurements under the current conditions.

In addition to analyzing the impact of lateral heat flow on $\lambda_S^*$, we also consider the effect of the Pt absorption pads that are along the path between the absorber under illumination and the spin detector. Since Pt is a well-known spin absorber [24], these unused pads will act as magnon sinks, thereby reducing the number of spins that reach the spin detector, and reducing $\lambda_S^*$ from its intrinsic value $\lambda_S$. A quantitative 2D FEM analysis of the magnon spin diffusion process in the presence of spin sinking indicates that $47 < \lambda_S < 73 \, \mu m$ (Appendix D).

We now compare the diffusion of magnon spins in YIG with the diffusion of magnon heat. The mean free path of magnons in bulk YIG was recently calculated to be 4 μm at 20 K [25]. Note that this mean free path reflects that of the magnons responsible for thermal transport, i.e. those magnons that contribute to the thermal conductivity of YIG. Therefore the mean free path reflects the distance between inelastic scattering events. However, our measurements demonstrate that at the same temperature, magnon spins diffuse more than one order of magnitude larger distances ($47 < \lambda_S < 73\ \mu m$).

Two explanations are offered for this observation. First, magnons retain their spin even after multiple inelastic scattering events. In that view, the long range magnons being detected are inelastically scattered magnons that still carry spin. This simple picture mirrors the case for electron spins in solids, where the spin-flip rate of electrons is typically far slower than the charge scattering rate. Thus, like electrons, magnons retain their spin polarization between successive linear momentum scattering events. In the second view, during thermal excitation of YIG, the thermal magnons that carry the majority of the heat scatter at short distances due to their short wavelengths and stronger interaction with the lattice. However, a small subset of the magnons, which are referred to as sub-thermal magnons and that are at much lower energies and have longer wavelengths, do not couple effectively to the lattice and therefore exhibit much longer mean free paths. As recently suggested [25], this can explain the ~100 nm length scale of longitudinal SSE at room temperature [14], the high magnetic field suppression of spin-Seebeck at room temperature [22,26] and is also consistent with our current observations at low temperature.

The long range diffusion of magnon spins at 23 K ($47 < \lambda_S < 73\ \mu m$), is larger than what was recently reported at room temperature by Cornelissen *et al*. (9 μm) [19], which indicates a temperature dependent inelastic scattering process for magnons in agreement with Boona and Heremans [25]. It is interesting to note the surprising similarity between the low temperature diffusion length of thermally-excited magnon spins in YIG compared to the room temperature coherence length of microwave spin-waves (Pirro *et al*.) [6]. Clearly additional measurements focusing on the temperature variation of magnon spin diffusion are required to understand the scattering processes further and connect the fields of microwave and thermal spintronics.

Since thermally-induced magnon spins in YIG are preserved over 100 μm, they may be used in laterally-patterned spin-current based devices powered by waste-heat. The low temperature spin diffusion length of YIG is far longer than that of n-GaAs (~6 μm at 4.2 K) [27]. It is worthwhile to consider using magnon spin conductors as the lateral spin channels (spin interconnects) in existing spin-based device proposals [28].

APPENDIX A: THERMAL MODELING

*I. Thermal Transport Modeling*

Thermal modeling is carried out using the commercially available 3D finite element modeling (FEM) software COMSOL [29]. Three dimensional steady-state heat diffusion equations are solved in Pt/YIG bi-layer structures using FEM. In the model, the YIG is 5 mm × 5 mm × 0.5 mm. The Pt spin detector is 265 μm × 265 μm × 10 nm (center region) and the absorption pads are 10 μm × 10 μm × 10 nm. All geometrical parameters match those of the device used in the non-local measurement. The bottom surface of the YIG is fixed at ambient temperature, mimicking the copper heat sink used in the measurement, while all other surface boundaries are thermally insulating, representing the vacuum in the cryostat. The laser is modeled as a Gaussian beam with a wavelength of 715 nm, and spot size of $1/e^2$ radius r = 3.75 μm. The spot size of the laser was acquired by a knife edge measurement [30].

The characteristics of the modeled laser match those of the laser employed in the measurement (width of 150 fs, 80 MHz repetition rate, modulated at a frequency of ~2 kHz). As ultrafast modeling (not shown here) indicates that a quasi-steady state is reached between each ultrafast pulse, we modeled the laser beam to consist of a square-wave (in reality a train of pulses) modulated at 2 kHz. The laser is absorbed in the center of the detector and is incident on the Pt between 0.614 and 0.886 ms. A smoothing zone of 0.025 ms is used for the rising and falling of the laser power corresponding to a rise time of 12.5 μs. The dark blue circles and light blue rectangles are the temporal response of the temperature along the optical axis at the Pt/YIG interface and 10 μm beneath the interface, respectively. In both cases, the temperature rapidly saturates to the steady-state solution with a rise time (10%~90%) of ~12.6 μs and ~23.5 μs, respectively. The resulting time profile of the variation of temperature of the Pt/YIG bilayer system under laser excitation is shown in Fig. 5. Given that the system rapidly approaches steady-state and the negligible residual excess temperature, we solve the time-independent heat diffusion equations under steady-state conditions for all of the modeling presented. Separate simulations are carried out for various average laser powers, including 10.5, 21.3 and 29.9 mW.

The reflectivity (R) and absorption coefficient (α) of Pt and YIG are calculated from the complex dielectric constant $\varepsilon = \varepsilon_1 + i\varepsilon_2$ at 23 K (300 K), where, for Pt [31], $\varepsilon_1$ = -16.26 (-14.36) and $\varepsilon_2$ = 16.86 (23.708) and for YIG [32], $\varepsilon_1$ ==4.28 (5.55) and $\varepsilon_2$ ==0 (0) at a wavelength of 715 nm. The values of $R$ and $\alpha$ are found to be 73.2% (69.5%) and $78.3 \times 10^6 \, m^{-1}$ ($80.1 \times 10^6 \, m^{-1}$) for Pt, and 12.1% (16.4%) and 0 (0) for YIG, respectively. We are only aware of low temperature optical data for Pt acquired at 77 K, and 87 K for YIG. As $R$ and $\alpha$ for Pt and YIG do not vary significantly at low temperature (77 K and 87 K) [33,34] and room temperature (300 K) [35] we expect that $R$ and $\alpha$ for Pt and YIG at 23 K are similar to their values at 77 K. Due to the thin nature of the Pt layer, we also considered the reflected power from the Pt/YIG interface that is being re-absorbed in the Pt layer. The reflectivity of the interface is calculated using the Fresnel equation [35] yielding $R_{interface}$ = 56% (46%). The thermal properties of Pt and YIG are acquired at 23 K and 300 K while the mass density data is measured at 300 K. All physical parameters used in the simulations are listed in Table I.

## II. Lateral heat flow

The simulated $\nabla T_x$ resulting from a heating laser with an average power of 10.5 mW, at the Pt/YIG interface, for 23 K and 300 K, are compared in Fig. 6. In this simulation, the laser is incident on 10 µm wide Pt absorption pad at a position of 60 µm from the edge of the Pt detector. $\nabla T_x$ is greater at 300 K due to the decreased thermal conductivity of Pt and YIG. Although the magnitudes are different, the decay length of $\nabla T_x$ is relatively unchanged at the two temperatures modeled, which is clearly illustrated by comparing the two dimensional contour map of $\nabla T_x$ shown in Fig. 6(b) to that shown in Fig. 1(c).

*III. Thermal loss analysis*

Heat loss due to convection cooling ($Q_{conv}^{loss}$) and black body radiation ($Q_{rad}^{loss}$) at the boundary are estimated using the Stefan-Boltzmann law and Newton's law [36],

$$Q_{conv}^{loss} = h * A * (T_{sur} - T_{amb}) \tag{1}$$

$$Q_{rad}^{loss} = \varepsilon * \sigma * A * (T_{sur}^4 - T_{amb}^4) \tag{2}$$

where $h$ is the heat transfer coefficient, $A$ is the sample surface area, $\varepsilon$ is the emissivity, $\sigma$ is the Stefan-Boltzmann constant, and $T_{sur}$ and $T_{amb}$ are the temperature of the sample surface and ambient, respectively. For the heat conduction loss, as all the measurements are conducted within a cryostat with a pressure below $4 \times 10^{-7}\ torr$, the heat transfer coefficient in this case is $h = 0\ W/m^2 K$ (for vacuum) [37] for which the convection loss $Q_{conv}^{loss} = 0$. For the radiation loss, using $\sigma = 5.67 \times 10^{-8}\ W/m^2 K^4$, $\varepsilon_{Pt} = 7.3 \times 10^{-4}$, $A = 10 \times 10\ \mu m$ and $T_{sur} = 30\ K$ (higher than the actual surface temperature on sample to estimate the upper bound of $Q_{rad}^{loss}$, actual $T_{sur}$ is ~28 K) [37,38] we find $Q_{rad}^{loss}$ ~$10^{-12}\ mW$, which is negligible in comparison to the laser power (30 $mW$). Heat loss due to convection cooling and black body radiation is therefore safe to neglect in the simulations. Besides convection and radiation losses, an additional heat loss mechanism is thermal conduction along the two voltage leads (annealed Au wire) from the sample surface and is estimated by Fourier's law of thermal conduction [36],

$$Q_{cond}^{loss} = \kappa * A * \nabla T \tag{3}$$

where $\kappa$ is the thermal conductivity and $\nabla T$ is the temperature gradient between two ends of the wire. Using $\kappa_{Au} = 11.4\ W/cmK$, $A = \pi * \left(\frac{25.4\ \mu m}{2}\right)^2 = 506.7\ \mu m^2$, and $\nabla T = \frac{23.015K - 23K}{1\ cm} = 0.015\ K/cm$, we find $Q_{cond}^{loss} = 1.7 \times 10^{-3}\ mW$ which is ~ 0.005 % of the absorbed laser power at 30 mW [39]. Here, 23.015K is the maximum temperature under the wire that is ~ 300 $\mu m$ away from the laser spot. Note that $Q_{cond}^{loss}$ is overestimated since the wire is 25.4 $\mu m$ in diameter and the temperature rise in the other contact region (between Au wire and Pt surface) is smaller than 0.015 $K$ making the average $\nabla T$ smaller than $0.015\frac{K}{cm} = 1.5 \times 10^{-6}\frac{K}{\mu m}$. However, even with the exaggerated value, $Q_{cond}^{loss}$ is still a small value compared to the total input power, therefore the heat loss due to the heat conduction through the wires is also neglected in the simulation, and we employ thermally insulating boundary conditions at the sample surface.

*IV. Estimation of Longitudinal SSE due to $\nabla T_z$*

When the laser is incident directly on the Pt detector, in the local configuration, an electric field is produced in the Pt beneath the laser due to the longitudinal spin-Seebeck effect (LSSE). Since the laser power follows a Gaussian distribution, the $\nabla T_z$ beneath the laser is not uniform as it is in the conventional LSSE measurement. Therefore, it is not appropriate to use a single value of $\nabla T_z$ to predict a measureable transverse voltage in the detection pad ($V_y$).

In the local measurement configuration, this non-uniformity of the laser induced $\nabla T_z$ leads to a spatially-dependent electric field in the Pt, $E_{ISHE}$, which decreases with increasing distance from the laser spot. The electric field is terminated in the region where the measurement wire is attached, due to the equi-potential nature of the wire. To account for the non-uniformity of $\nabla T_z$ beneath the laser, we integrate the FEM predicted values of $\nabla T_z$ at Pt/YIG interface, along a line through the center of the region illuminated by the laser ($\int \nabla T_z \, dl$). We find that $\int \nabla T_z \, dl$ exhibits the same power dependence as $V_y$, as shown in Fig. 7. The slope is different only in magnitude, implying that this integral, $\int \nabla T_z \, dl$ is proportional to the measured transverse voltage across the detection pad, $V_y$.

This allows us to determine a linear relation between the thermal gradients beneath the detection pad and the measured voltages. We plot $V_y$ as a function of $\int \nabla T_z \, dl$ and thereby determine the effective longitudinal spin-Seebeck effect coefficient ($\alpha_{LSSE}$) of our local measurement configuration. This is found to be ~102 nV/K, as shown in Fig. 8. Table II provides a summary of the values previously described, including the FEM modelled values of $\nabla T_{z\_max}$ (the maximum thermal gradient at the Pt/YIG interface beneath the heating laser) and the calculated values of $\int \nabla T_z \, dl$ for lines through the center of for regions beneath the heating laser) for three different laser powers in the local configuration. $V_y$ at each power is also included.

We can now use these results to determine how large a spurious $\nabla T_z$ beneath the detection pad would have to be to cause the observed $V_y$ in our non-local measurements. To do this, we consider a specific non-local measurement in which the laser is focused on an absorption pad 54 μm from the detection pad. The measured $V_y$ is 25 nV. From linear extrapolation of the $V_y$ vs. $\int \nabla T_z \, dl$ graph, it is found that in order to produce a $V_y$ value of 25 nV a value of $\int \nabla T_z \, dl = 0.34$K is needed beneath the detection pad (See Fig. 8). As previously discussed, radiative losses and conductive losses through the measurement wires attached to the detection pad (12.7 μm radius) result in at most a $\nabla T_z$ $1.5 \times 10^{-6} \frac{K}{\mu m}$ to flow through the detection pad. This would yield a $V_y$ of at most, 0.5 pV. It is illustrative to calculate the diameter of detection wires that would be needed to account for the measured non-local voltage. As previously stated, this would require $\int \nabla T_z \, dl$ = 0.34 K. We find that wires of 200 μm radii (0.4 mm diameter) are necessary and that a temperature difference across the length of the wires of $\Delta T_{wire} = T_{cryostat} - T_{Pt} = 168$ K would be required to allow for the required heat flow through the Pt detection strip. Given that the actual wires utilized are more than one order of magnitude smaller in diameter (>100× smaller cross-sectional area and thermal conductance), and that an unphysical platinum temperature, $T_{Pt}$= -188 K, would be required to obtain $\Delta T_{wire} = 168$ K ($T_{cryostat} = 20$ K), then $\nabla T_z$ cannot contribute to the non-local signal.

APPENDIX B: IMPACT OF LATERAL HEAT FLOW ON MAGNON SPIN DIFFUSION

A single decaying exponential is fit to multiple $V_{NL}$ data sets containing a varying number of data points. Fits were constructed from data sets that include Δx values > 23 μm, where FEM modeling predicts that $\nabla T_x$ is negligible (defined as zone II), and also from data sets with 0 < Δx < 23 μm, where $\nabla T_x$ may not be negligible (defined as zone I). As can be seen from Tables III and IV, there is no significant variation in the $R^2$ value between fits including data points from both zones I and II and those containing data points exclusively from zone II. In addition, Fig. 9 and Fig. 10 show that the fits lie within the error bars for single exponential decaying functions fit to $V_{NL}$ values from both zones I and II. This indicates that $\nabla T_x$ shows negligible effect on magnon spin diffusion under the experimental conditions tested.

APPENDIX C: MATERIALS, METHODS, AND MEASUREMENT SYSTEMATICS

*I. Sample processing and characterization*

Single-side-polished single-crystal <100> YIG samples with dimensions of 5 × 5 × 0.5 mm$^3$ are obtained commercially from the Princeton Scientific Corporation. The polished surface of the YIG is atomically flat and epi-ready with a surface roughness of 0.27 nm measured by atomic force microscopy (AFM) using a Bruker Dimension Icon AFM tool. The 10 nm Pt is e-beam evaporated on the YIG at a rate of $0.15\ \dot{A}/s$ using a Kurt Lesker LAB 18 deposition system with a base pressure $< 2.5 \times 10^{-7}$ torr. Prior to the Pt deposition, the YIG samples are cleaned using solvent and DI water, followed by a five-minute dehydration bake at 150 °C in atmosphere and a second one-hour in-situ bake at 150 °C. There is no vacuum break between the in-situ baking and the actual Pt deposition in order to ensure a high quality Pt/YIG interface. The Pt is patterned into a $265 \times 265\ \mu m$ (center region) detection pad and $10 \times 10\ \mu m$ absorbtion pads with $3\ \mu m$ spacing using standard photo-lithography and $Cl_2/CF_4$ based reactive-ion etching (RIE). The magnetic properties of the YIG sample are measured by superconducting quantum interference device (SQUID) magnetometry using a Quantum Design MPMS XL. The saturation magnetization is $220\ emu/cm^3$ and the coercivity is $5\ Oe$ with an in-plane magnetic field (same orientation as SSE measurement) at 20 K.

## II. Opto-thermal spin-Seebeck measurement

The Pt detector is wired with two $25.4\ \mu m$ diameter gold wires and the YIG is attached to a copper block heat sink using silver paste. All measurements are conducted in high vacuum in a customized closed cycle He cryostat coupled to an electromagnet capable of producing magnetic fields of $\pm\ 0.25$ T at the sample. A Chameleon Ultra II Ti:Sapphire laser is focused through a reflective microscope objective, producing a 7.5 μm diameter laser spot on the sample surface. Mechanically chopping the laser at 2 kHz to serve as a reference frequency allows the induced voltage to be measured using standard lock-in technique. Error bars are calculated based on the standard deviation of the value $V_L$ and $V_{NL}$ extracted from the magnetic field dependent measurements as defined in the text.

III. *Measurement systematics*

Additional non-local measurements are carried out for the laser positioned to the left and the right side of the Pt spin detector and the results are shown in Fig. 9 and Fig. 10. With only one outlying point (59 µm), these systematic measurements reveal almost identical values of $\lambda_S^*$ (~47 µm) showing the lack of a laser power dependence on left/right asymmetry, as well as confirming the reproducibility of the experiment.

Non-local measurements are also carried out on a different Pt/YIG sample (20141004). As shown in Fig. 11, both $V_L$ and $V_{NL}$ track the magnetization. When the laser is focused on the detection pad, $V_L$ is ~270 nV, while when the laser is 47 µm away, $V_{NL}$ is approximately 9 nV. As the laser is moved away from the detection pad, $V_{NL}$ falls off exponentially, as expected. A single decaying exponential fits the data with $V_{NL}$ from both zones I and II with an adjusted $R^2$ value of .998 (Table V). $\lambda_S^*$ is constant regardless of the relative laser position (left side versus right side of the spin detector) shown in Fig. 12.

APPENDIX D: EFFECT OF SPIN SINKING AND THE UPPER BOUND OF MAGNON SPIN DIFFUSION LENGTH

As previously mentioned, it is well known that Pt acts as a strong spin absorber. However, the measured value of $\lambda_S^* = 47\ \mu m$ does not take this spin sinking into account. Therefore it serves as a lower bound for $\lambda_S$. It is important to also establish an upper bound for $\lambda_S$ by taking into account the loss of magnons en route to the detector via spin sinking from the unused Pt absorption pads. To do this we perform 2D FEM to solve the diffusion equation for nonequilibrium magnons in the presence of Pt spin sinks. The magnon diffusion is specified by [40]

$$D\nabla^2 \delta m_m - \frac{\delta m_m}{\tau_{th}} = 0 \qquad (4)$$

where $D$ is the magnon diffusivity in YIG, $\delta m_m$ is the magnetic moment density (magnon density times $2\mu_B$), and $\tau_{th}$ is the magnon lifetime. The magnon flux from YIG into a Pt spin sink is described by

$$-\hat{n} \cdot (-D\nabla \delta m_m) = -G_{me}\delta m_m \tag{5}$$

where $\hat{n}$ is a unit vector normal to the YIG surface and $G_{me}$ is the spin convertance. $G_{me}$ is estimated as [41]

$$G_{me} \sim \frac{\pi S a_{0I}^5 J_{sd}^2 g_e(\varepsilon_F) T}{\hbar a_{0M} T_F} \tag{6}$$

where $a_{0I}$ and $a_{0M}$ are the lattice constants of the YIG and the Pt respectively, $J_{sd}$ is the exchange interaction, $g_e(\varepsilon_F)$ is the Pt density of states at the Fermi energy and $T_F$ is the Fermi temperature of Pt. Using $a_{0I} = 12.376$ Å, $a_{0M} = 3.900$ Å, $J_{sd} = 1.000\ meV$, $g_e(\varepsilon_F) = 1.164 \cdot 10^{23}\ states/(eV \cdot cm^3)$ [42], and $T_F = 9.812 \cdot 10^4$ [42] $G_{me}$ is calculated to be 31.598 m/s. Note that although these parameters are reported at 300K, they are expected to be relatively temperature independent. This estimate of $G_{me}$ serves as a theoretical upper bound on the degree of spin sinking from YIG into Pt. From reference [41], we utilize $\tau_{th} = 10^{-6}$ s.

Following the calculation of $G_{me}$, eqns. (3) and (4) are simultaneously solved in order to calculate the magnon density profile during the non-local measurements. The geometry of the FEM is identical to the experimental setup and consists of a 265 μm spin detector centered on a 5 mm wide YIG crystal that is 500 μm thick. The right most edge of the detector is at x = 0. The 10 μm wide Pt absorption pads are spaced 3 μm apart, with each pad further from the edge of the detector. The boundary conditions force $\delta m_m$ = 0 at the edges of the YIG crystal, a few mm from the point of magnon excitation.

Results of the FEM modeling are shown in Fig. 13a, where a 2D map of the steady state $\delta m_m$ distribution is plotted. In this example, the laser is incident in the center of the Pt absorber at Δx = 47 µm. The interfacial magnon density profile is plotted by slicing the full 2D profile along Δx at z = 0, i.e. at the Pt/YIG interface, and shown in Fig. 13b. Three cases are shown to illustrate the effect of spin sinking on the magnon density profile. As expected, spin sinking reduces the overall magnon density.

The derivative of the magnon density profile, $\delta m_m$, along the z-direction evaluated at z = 0, $\nabla_z \delta m_m$, is plotted in Fig. 13c along Δx, which shows the magnon concentration gradient responsible for the vertical diffusion of magnons into the absorbers and spin detector. To determine the total magnon spin current flowing into the detector, $J_s^z$, we calculate the average of $\nabla_z \delta m_m$ at z = 0 over the entire detector width along x and multiply by $-D$, i.e. $J_s^z = -\frac{D}{265} \int_{\Delta x = -265}^{0} \nabla_z \delta m_m \, d\Delta x$. The above procedure is carried out with the magnon injection point being varied along each of the various detector pads (just as in the experiment) and $J_s^z$ is determined under each of these conditions. The results are plotted in Fig. 13d.

The value of $D$ is treated as a fit parameter and adjusted to achieve $\lambda_S^* = 47$ µm, in order to match the experimental measurements to the FEM model (dashed orange line in Fig. 13d). From this fit we determine a magnon diffusivity of $D = 0.0053$ m²/s and therefore since $\lambda_s = \sqrt{D \tau_{th}}$ we find $\lambda_s = 73 \, \mu m$. Because the theoretical maximum value of $G_{me}$ and $\tau_{th}$ were used in this model, it defines the upper bound for $\lambda_s$. Therefore, $47 < \lambda_S < 73 \, \mu m$.

$G_{me}$, which parameterizes the spin sinking ability of the Pt absorbers, was estimated using the largest reasonable parameters from the literature. As such, it is the theoretical maximum. The other parameter that will have the largest impact on the numerical solution to eqns (3) and (4) is $\tau_{th}$. To determine how sensitive our model is to this parameter, a value of $\tau_{th} = 10^{-8}$ s was also chosen as an input parameter. Using this value, a fit parameter of $D = 0.41$ m²/s is found, yielding $\lambda_s$ = 64 $\mu m$. Even by varying $\tau_{th}$ two orders of magnitude, the calculated value for $\lambda_s$ remains relatively constant.

## ACKNOWLEDGEMENTS

The authors thank J. P. Heremans and S. R. Boona for useful discussions. This work was primarily supported by the Army Research Office MURI W911NF-14-1-0016. J.J. acknowledges the Center for Emergent Materials at The Ohio State University, an NSF MRSEC (Award Number DMR-1420451), for providing partial funding for this research.

FIGURE CAPTIONS

FIG. 1. (a) Schematic diagram of the non-local spin current injection/detection geometry (not to scale). (b), (c) Three-dimensional FEM thermal modeling was carried out with the laser positioned over an absorption pad at $\Delta x = 60$ μm, indicating spherically symmetric heat flow. Two-dimensional color maps of $\nabla T$, directly below the absorption pad and in the x-z plane, are presented. $\nabla T$ in the vertical direction, ($\nabla T_z$) is shown in (b) and $\nabla T$ in the lateral direction ($\nabla T_x$) is shown in (c). Temperature gradients are plotted as a function of distance from the edge of the spin detector ($\Delta x$) and as a function of depth from the surface of YIG (z). The edges of the absorption pad are indicated by dashed lines in revealing that both $\nabla T_x$ and $\nabla T_z$ are well isolated from the edge of the spin detector (at $\Delta x = 0$ μm). The color scale is logarithmic.

FIG. 2. (a) Schematic of the local detection geometry. (b) The transverse voltage across the spin detector ($V_y$) as a function of the applied in-plane magnetic field ($B_x$) with the laser positioned on the spin detector as shown in (a). The measured magnetization ($M$) is included for comparison. The magnitude of the local voltage is defined as $V_L$, as shown in (b) and may contain components not purely due to spin currents. The inset plots $V_L$ as a function of laser power. (c) Schematic of the non-local spin-Seebeck geometry. (d) $V_y$ as a function of $B_x$ with the laser positioned on a remote absorption pad as shown in (c). $M$ is also plotted for comparison. The magnitude of the non-local signal, $V_{NL}$, is proportional to the magnon spin current diffusing from the absorber to the detector, which is plotted in the inset as a function of laser power.

FIG. 3. (a) Measurement of $V_{NL}$ as the laser is focused on absorption pads of increasing distance at 23 K (black) compared to FEM in-plane spurious temperature gradients at the edge of the detection pad (blue) at 23K. (b) Comparison of same measurement/prediction as (a) but at 280K. As can be seen, even though FEM predicts higher spurious temperature gradients at 280K $V_{NL}$ becomes negligible, proving that spurious temperature gradients cannot be the cause of the signal measured at 23K.

FIG. 4. (a) The magnitude of the non-local signal $V_{NL}$ as a function of laser position from the edge of the spin detector ($\Delta x$) measured with three different laser powers. The signal reveals exponential decay of the magnon spin current produced at the absorber and diffusing to the spin detector. $\nabla T_x$ is included for comparison (dashed lines are guides to the eye). The $V_{NL}$ data are well fit to a single decaying exponential (solid lines) to obtain the pure magnon spin diffusion length ($\lambda_S^*$) plotted in (b) as a function of laser power.

FIG. 5. The temporal response of the temperature along the optical axis after having been heated by the laser pulse train modulated at 2 kHz, with a fluence of $3.2 \times 10^{16} \frac{W}{m^3}$. Dark blue circles indicate the response at the Pt/YIG interface, while light blue rectangles indicate the response 10 μm below the interface. The red dashed line indicates the power of the laser at the center of the optic axis. The blue dashed lines represent calculated steady state values of the temperature rise.

FIG. 6. (a) A comparison of the one dimensional $\nabla T_x$ profile at the Pt/YIG interface at 23 K and 300 K. The laser is focused 60 μm away from the detector. (b) Two dimensional temperature gradient contour map of $\nabla T_x$ (cross-section view) with the laser absorbed 60 μm away from the detector at 300 K. The average laser power used for the simulation was 10.5 mW.

FIG. 7. A comparison of $\int \nabla T_z \, dl$ and $V_y$ as a function of laser power shows the slopes are nearly equal, implying that $V_y \propto \int \nabla T_z \, dl$. $\nabla T_z$ is integrated in the y-direction at the Pt/YIG interface, across the center of the heating laser.

FIG. 8. The linear fit (red dotted line) between $V_y$ (black dots) and $\int \nabla T_z \, dl$. An LSSE coefficient of 102 $nV/K$ is extracted from the slope. The blue spot corresponds to a $V_y$ of 25 nV. From the linear relationship, this requires a corresponding $\int \nabla T_z \, dl$ of 0.34K. This implies $\nabla T_z$ is negligible beneath the detection pad.

FIG. 9. An analysis of $V_{NL}$ for multiple different sets of absorption pads located on either side of the spin detector. The laser powers are 10.5 mW in (a), (b), 21.3 mW in (c), (d), and 29.9 mW in (e), (f). Green and red dotted lines correspond to the best fit single decaying exponential when the laser is to the left (x < 0) or right (x > 0) of the spin detector, respectively. Extracted spin diffusion lengths (in μm) are shown in the corresponding figure panels. All measurements were performed at 23 K. As expected, the extracted spin diffusion length does not depend on the set of absorption pads chosen.

FIG. 10. Data from Fig. 9 plotted on logarithmic scale in order to emphasize the validity of a single exponential fit. The fits lie within error bars for all data points.

FIG. 11. (a) $V_L$ is measured with the laser focused directly on the spin detector on sample 20141004. The signal tracks the magnetization, following the same trends as seen on sample

20140804. (b) $V_{NL}$ is measured on sample 20141004, with the laser on an absorption pad 47 μm away. As seen in sample 20140804, $V_{NL}$ is an order of magnitude smaller than $V_y$, and is attributed to diffusion of the magnon mediated spin current. Comparison of to Fig. 2 (sample 20140804) confirms that the non-local measurement is not sample dependent.

FIG. 12. Laser power is 10.5 mW. Green and red dotted lines correspond to the best fit single decaying exponential when the laser is to the left or right side of the spin detector, respectively. Extracted spin diffusion lengths (in μm) are shown in the corresponding figure panels. All measurements were performed at 24 K.

FIG. 13. (a) A 2D map of the steady state $\delta m_m$ distribution, depicting the experimental setup where the laser is focused on the Pt absorber at Δx = 47. (b) The interfacial magnon density profile along Δx at z = 0 (Pt/YIG interface). Three scenarios are depicted, showing the result with no spin sinking, when only the detector serves as a spin sink and when unused Pt absorbers act as spin sinks. (c) The derivative of the magnon density profile, $\delta m_m$, along the z-direction evaluated at z = 0, $\nabla_z \delta m_m$. The value of $\delta m_m$ under the detector is integrated to determine the total magnon spin current flowing into the detector, $J_s^z$. (d) The results of the integration described in (c) when the laser is focused on varying absorption pads. By varyng $\tau_{th}$ over two orders of magnitude the value of $\lambda_s$ remains relatively constant.

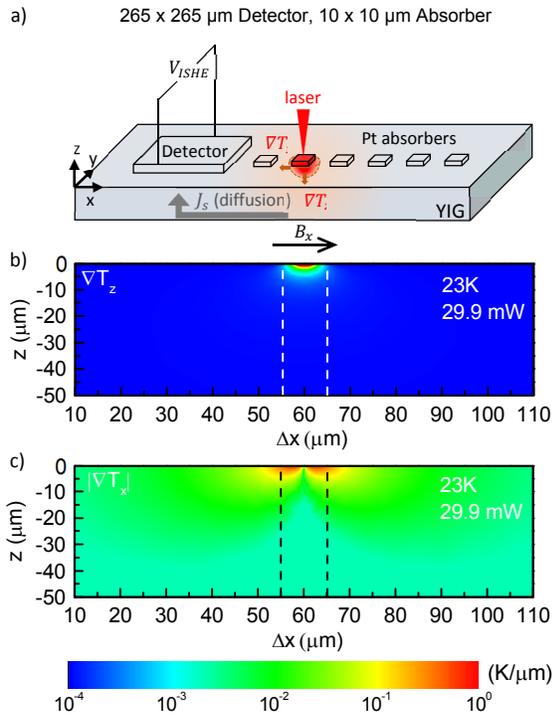

FIG. 1 (Single Column; Color online)

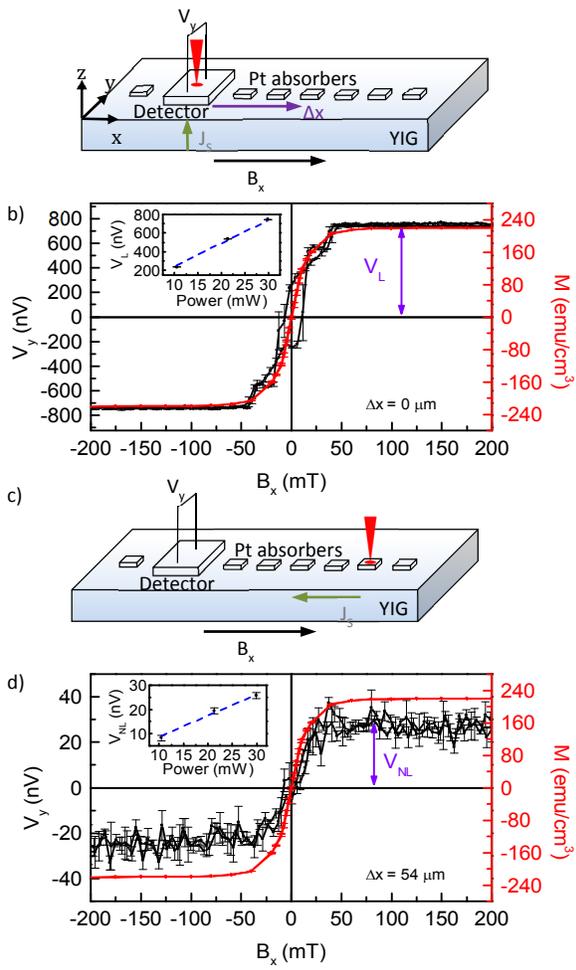

FIG. 2 (Single Column; Color online)

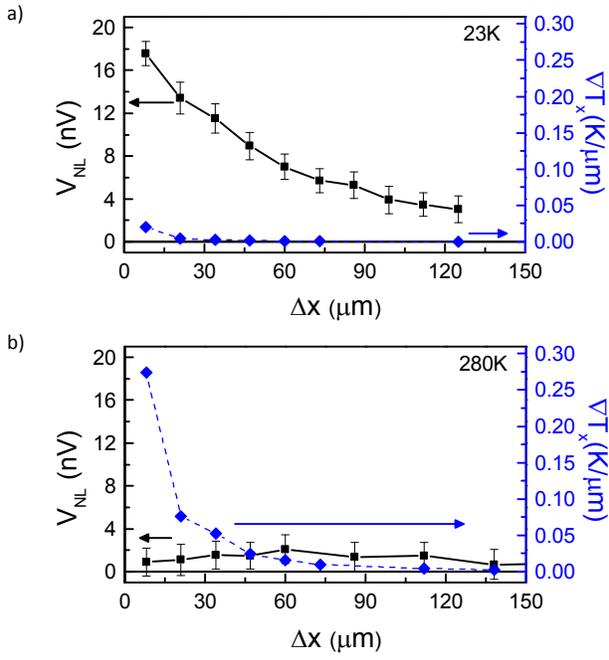

FIG. 3 (Single Column; Color online)

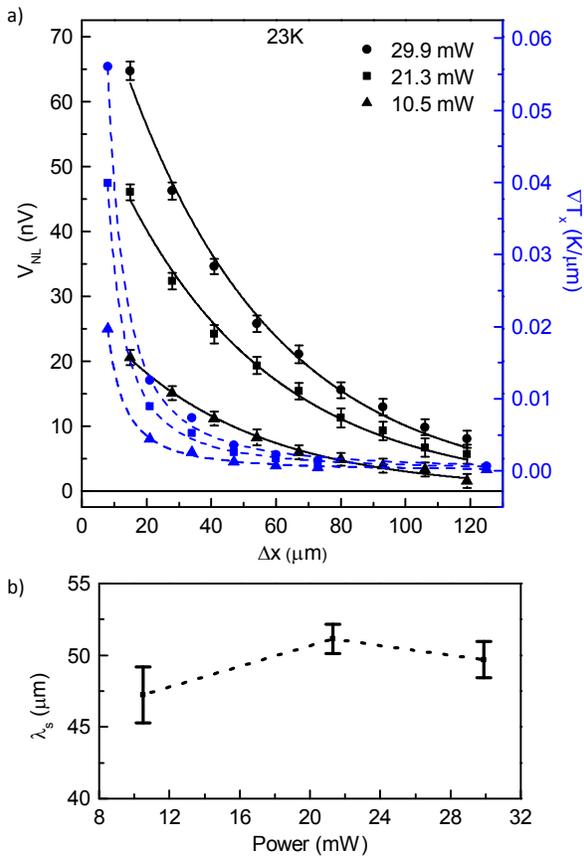

FIG. 4 (Single Column; Color online)

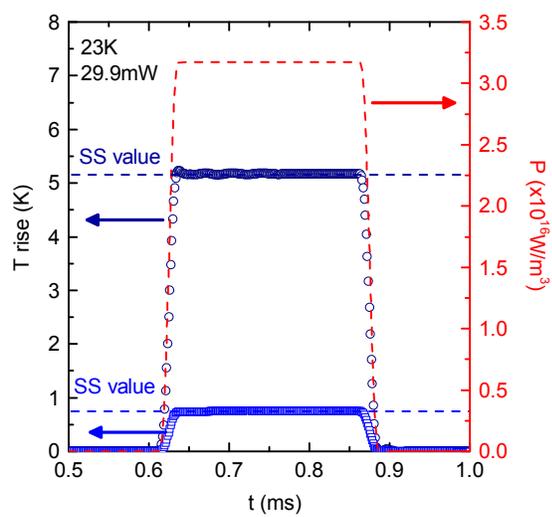

FIG. 5 (Single Column; Color online)

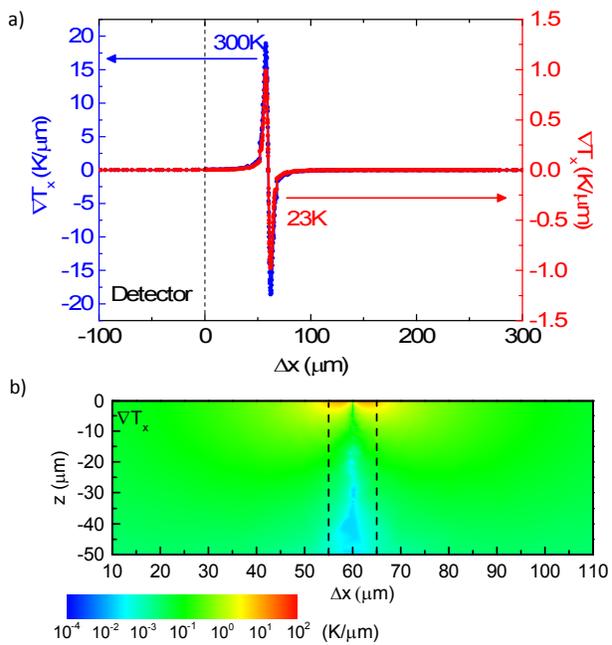

FIG. 6 (Single Column; Color online)

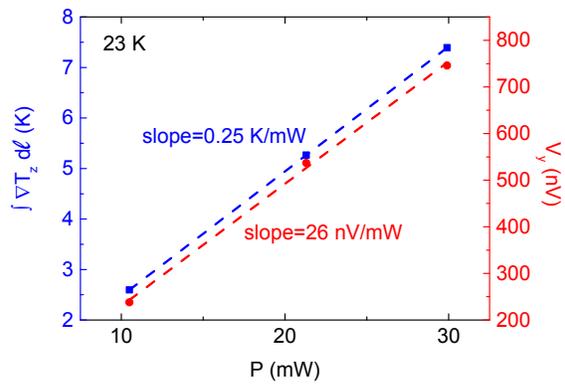

FIG. 7 (Single Column; Color online)

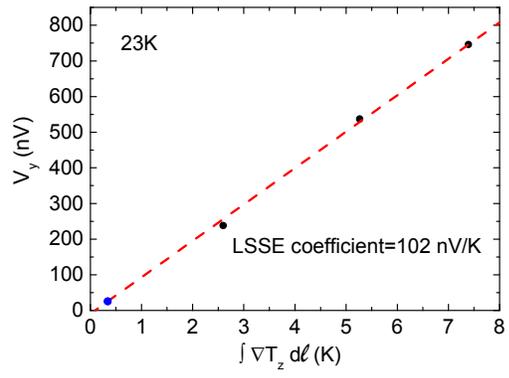

FIG. 8 (Single Column; Color online)

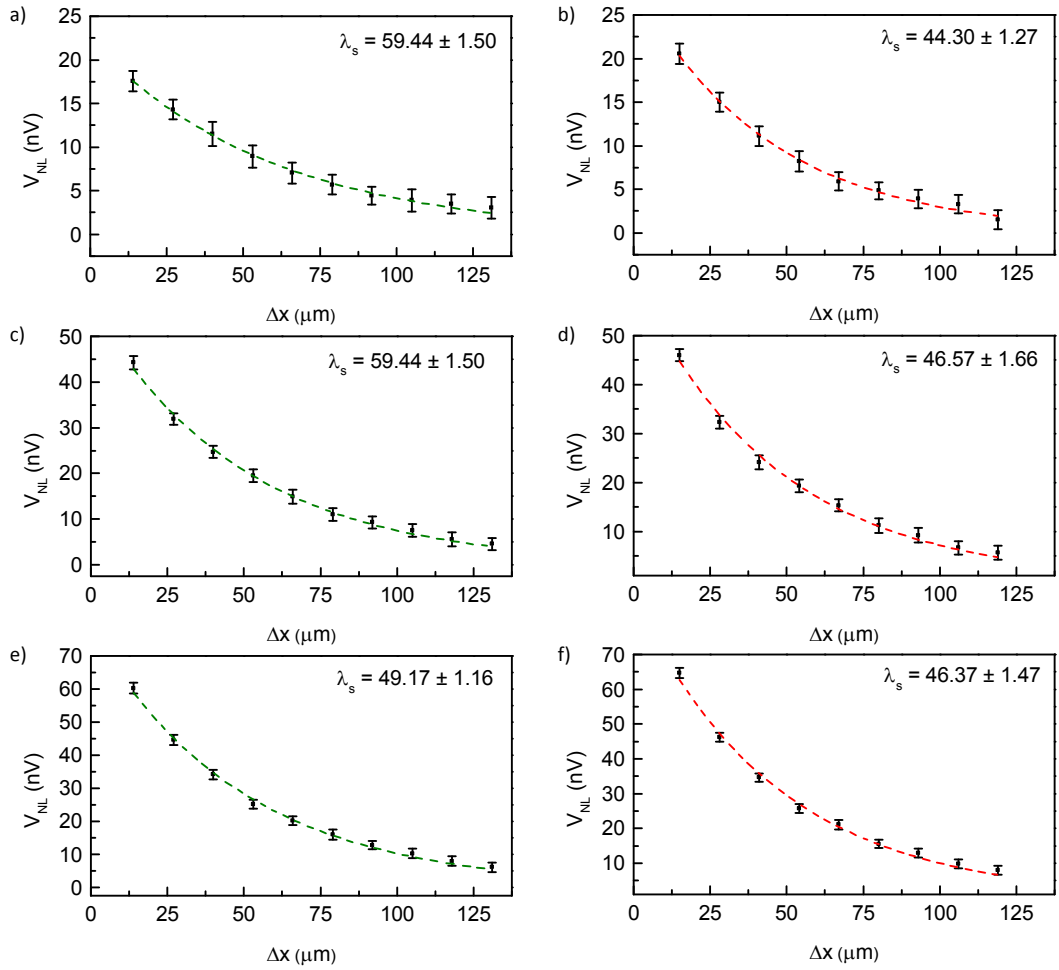

FIG. 9 (Double Column; Color online)

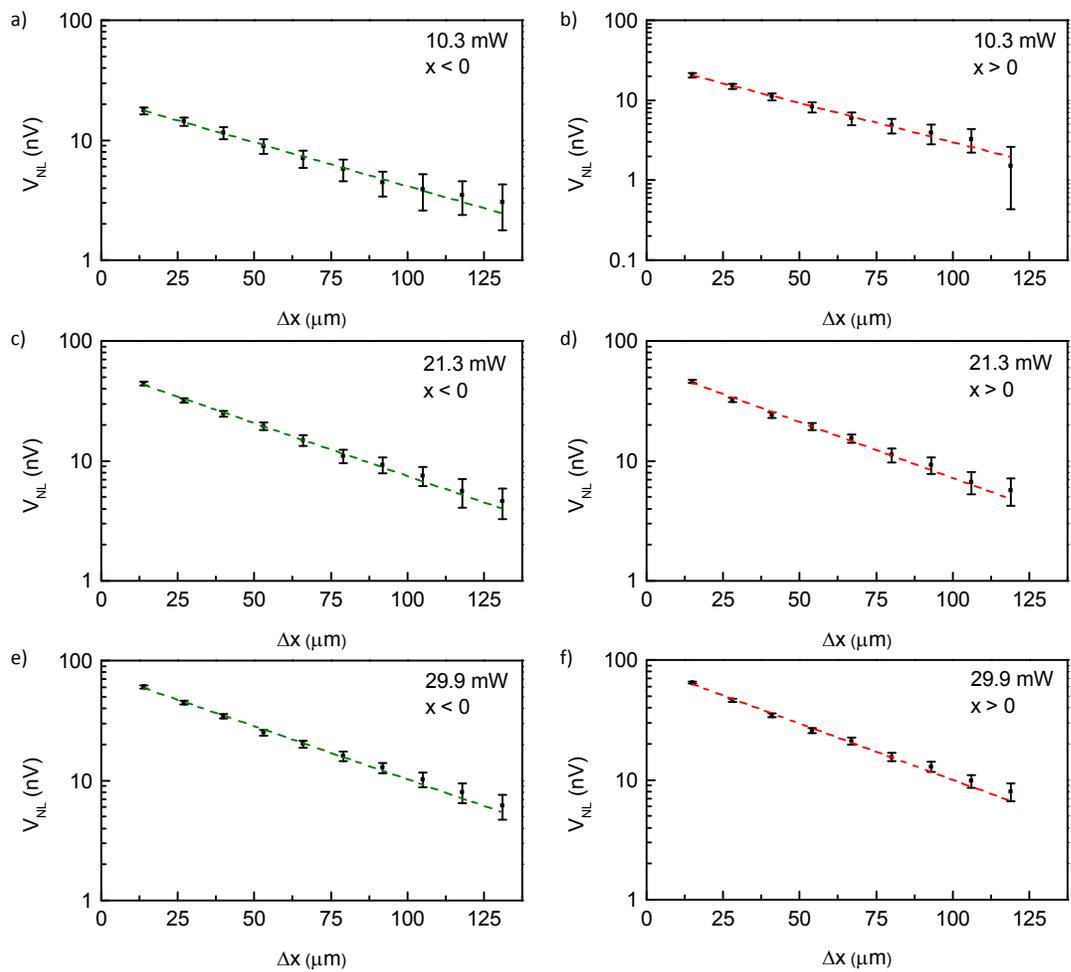

FIG. 10 (Double Column; Color online)

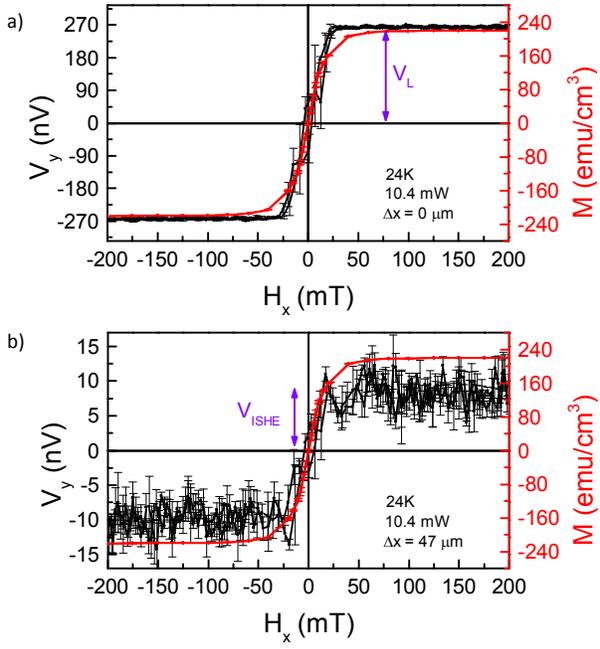

FIG. 11 (Single Column; Color online)

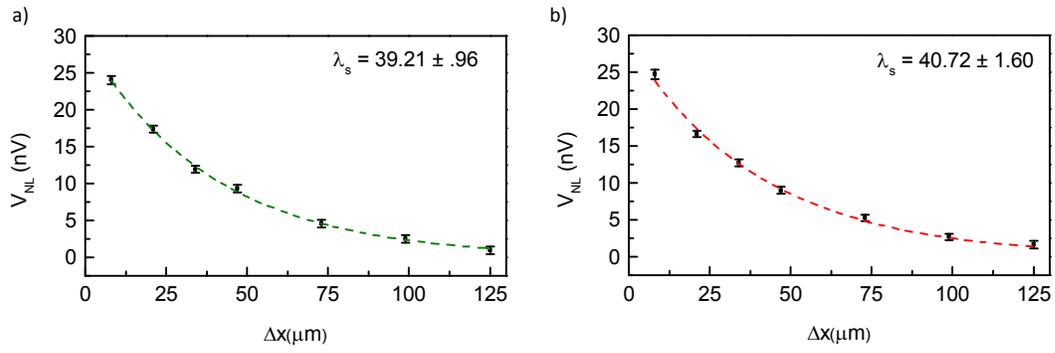

FIG. 12 (Double Column; Color online)

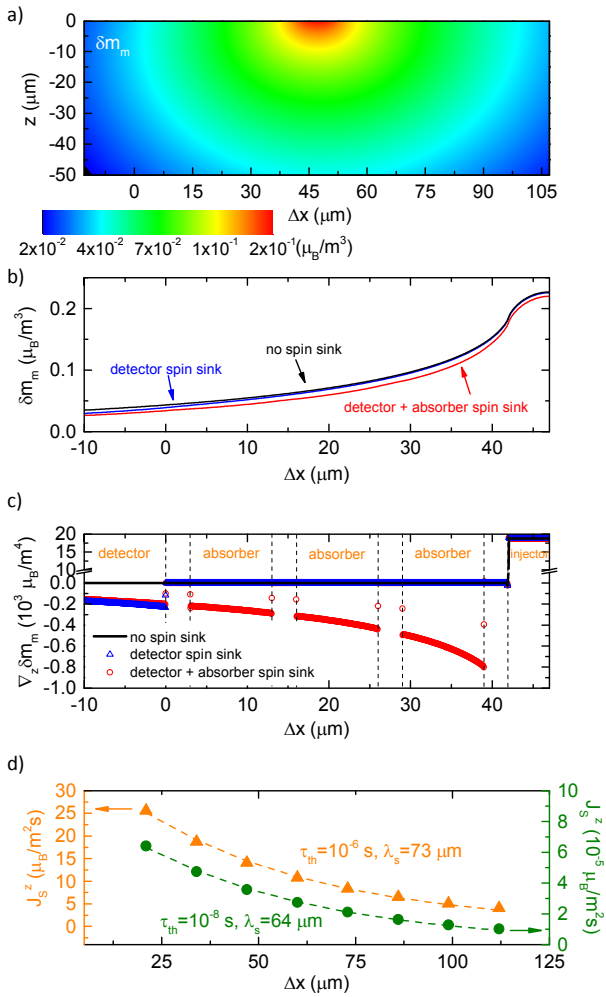

FIG. 13 (Single Column; Color online)

TABLE I. Refractive index ($n$), reflectivity ($R$), absorption coefficient ($\alpha$), density ($\rho$), thermal conductivity ($\kappa$) and heat capacity ($C$) at 23 K and 300K used in the simulation.

|     |      | n              | R (to air) | α (1/m)         | ρ (kg·m³) | κ (W·m⁻¹·K⁻¹) | C (J·kg⁻¹·K⁻¹) |
|-----|------|----------------|------------|-----------------|-----------|---------------|-----------------|
| Pt  | 23K  | 1.89 + 4.54i[a] | 73.2%     | 78.3 x 10⁶ [a]  | 21450[c]  | 350[d]        | 11.7[e]         |
|     | 300K | 2.58 + 4.59i[b] | 69.5%     | 80.1 x 10⁶ [b]  | 21450[c]  | 72[c]         | 130[c]          |
| YIG | 23K  | 2.07 + 0i[f]    | 12.1%     | 0               | 5245[c]   | 109.5[g]      | 5.6[g]          |
|     | 300K | 2.36 + 0i[f]    | 16.4%     | 0               | 5245[c]   | 6[c]          | 570[c]          |

[a] Reference [29].
[b] Reference [33].
[c] Reference [35].
[d] Reference [39].
[e] Reference [43].
[f] Reference [32].
[g] Reference [30].

TABLE I. (Double Column)

TABLE II. $\alpha_{LSSE}$, $\int \nabla T_z\, dl$, $\nabla T_{z\_max}$ and $V_y$ are shown as a function of laser power for local measurement.

| P (mW) | $\nabla T_{z\_max}$ (K/$\mu m$) | $\int \nabla T_z\, dl$ (K) | $V_y$ (nV) |
|---|---|---|---|
| 10.5 | 0.78 | 2.60 | 238.1 |
| 21.3 | 1.58 | 5.27 | 537.1 |
| 29.9 | 2.21 | 7.39 | 745.9 |

TABLE II. (Double Column)

TABLE III. Multiple data sets containing $V_{NL}$ values from both zones I and II and from exclusively from zone II (Sample 20140804). Values indicate no significant difference, indicating that $\nabla T_x$ can be considered negligible.

| | Left Side (x < 0) | | | Right Side (x > 0) | |
|---|---|---|---|---|---|
| Power (mW) | Absolute value of Δx (µm) | Adj $R^2$ | Power (mW) | Absolute value of Δx (µm) | Adj $R^2$ |
| 10.5 | 79 – 118 | 0.9323 | 10.5 | 80 – 119 | 0.99136 |
| | 66 – 118 | 0.96958 | | 67 - 119 | 0.99618 |
| | 53 – 118 | 0.98175 | | 54 – 119 | 0.97702 |
| | 40 – 118 | 0.98757 | | 41 – 119 | 0.98384 |
| | 27 – 118 | 0.99469 | | 28 – 119 | 0.99112 |
| | 14 – 131 | 0.99596 | | 15 – 119 | 0.99451 |
| 21.3 | 79 – 118 | 0.97598 | 21.3 | 80 – 119 | 0.97851 |
| | 66 – 118 | 0.98487 | | 67 - 119 | 0.99017 |
| | 53 – 118 | 0.99246 | | 54 – 119 | 0.99556 |
| | 40 – 118 | 0.99696 | | 41 – 119 | 0.99661 |
| | 27 – 118 | 0.99863 | | 28 – 119 | 0.99801 |
| | 14 – 131 | 0.99622 | | 15 – 119 | 0.99375 |
| 29.9 | 79 – 118 | 0.99882 | 29.9 | 80 – 119 | 0.99001 |
| | 66 – 118 | 0.99963 | | 67 - 119 | 0.98997 |
| | 53 – 118 | 0.99981 | | 54 – 119 | 0.99546 |
| | 40 – 118 | 0.9948 | | 41 – 119 | 0.99657 |
| | 27 – 118 | 0.99629 | | 28 – 119 | 0.99679 |
| | 14 – 131 | 0.99654 | | 15 – 119 | 0.99581 |

* Yellow regions refer to data sets with $V_{NL}$ values from zones I and II. Blue regions refer to data sets containing $V_{NL}$ values from zone II only.

TABLE III. (Double Column; Color online)

TABLE IV. Multiple data sets containing $V_{NL}$ values from both zones I and II and from exclusively from zone II (Sample 20141004). Values indicate no significant difference, indicating that $\nabla T_x$ can be considered negligible.

| Left Side (x < 0) | | | Right Side (x > 0) | | |
|---|---|---|---|---|---|
| Power (mW) | Absolute value of Δx (μm) | Adj $R^2$ | Power (mW) | Absolute value of Δx (μm) | Adj $R^2$ |
| 10.5 | 47 – 125 | 0.99724 | 10.5 | 47 – 125 | 0.99709 |
|  | 34 – 125 | 0.99749 |  | 34 – 125 | 0.99391 |
|  | 21 – 125 | 0.99876 |  | 21 – 125 | 0.99695 |
|  | 8 – 125 | 0.99545 |  | 8 – 125 | 0.99843 |

* Yellow regions refer to data sets with $V_{NL}$ values from zones I and I. Blue regions refer to data sets containing $V_{NL}$ values from zone II only.

TABLE IV. (Double Column; Color online)

TABLE V. Decaying single exponential fit parameters as a function of laser power and laser position relative to the spin detector.

| Power (mW) | Left Side (x < 0) | | | Right Side (x > 0) | | |
|---|---|---|---|---|---|---|
| | $V_0$ (nV) | $\lambda_s$ (μm) | Adj $R^2$ | $V_0$ (nV) | $\lambda_s$ (μm) | Adj $R^2$ |
| 10.5 | 22.2 ± .41 | 59.44 ± 1.50 | .996 | 28.5 ± .70 | 44.30 ± 1.27 | .996 |
| 21.3 | 57.1 ± 1.20 | 49.30 ± 1.30 | .996 | 61.2 ± 1.69 | 46.57 ± 1.66 | .994 |
| 29.9 | 78.9 ± 1.57 | 49.17 ± 1.16 | .997 | 86.7 ± 2.34 | 46.35 ± 1.47 | .995 |

*Sample 20140804

| Power (mW) | Left Side (x < 0) | | | Right Side (x > 0) | | |
|---|---|---|---|---|---|---|
| | $V_0$ (nV) | $\lambda_s$ (μm) | Adj $R^2$ | $V_0$ (nV) | $\lambda_s$ (μm) | Adj $R^2$ |
| 10.5 | 29.4 ± .48 | 39.21 ± .96 | .998 | 29.1 ± .80 | 40.72 ± 1.60 | .995 |

*Sample 20141004

TABLE V. (Double Column)